\documentclass[10pt,letterpaper]{article}

\usepackage[,margin=1.3in]{geometry}
\usepackage[latin1]{inputenc}
\usepackage{amsmath}
\usepackage{amsfonts}
\usepackage{amssymb}
\usepackage{makeidx}
\usepackage{graphicx}
\usepackage{hyperref}
\usepackage{enumitem}
\setlist{nosep}

\title{awkt: A Parameter Estimation Tool for Physicochemical Characterization }
\author{Christian Löffeld\footnote{Email: christian.loeffeld@gmail.com} }
\begin{document}
\maketitle

\section*{Abstract}

We present a computational parameter estimation tool aimed at assisting molecule characterization applying nonlinear least-squares optimization to a specific closed-form solution of a capillary electrophoretic transport problem. Numerical estimates for the respective physicochemical parameters are generated. Among other features, the tool facilitates automated and user-controllable hypothesis testing and model selection. It can also be used to perform fully automated batch-processing in true concurrent mode.
\section{Introduction}
Capillary zone electrophoresis is an electrophoretic separation technique that is utilized in a large variety of chemical and biochemical analysis settings \cite{Frost,Moritz,Ouimet}. More recently, electrophoretic methods have become important tools in proteomic and genomic endeavors, however most often coupled to a downstream mass spectrometry setup \cite{Han, Li}. Given appropriate experimental conditions, capillary zone electrophoresis has the potential to enable deep physicochemical analyses of complex analytes such as proteins on its own. The hydrodynamic radius and the net electric charge of a molecular species are of considerable interest with regards to its characterization. Capillary zone electrophoresis has an intimate handle on these two quantities \cite{Jorgenson,Grossman}. Reliable determination of them would transform CZE into a truly multidimensional analytical tool, i.e. simultaneously facilitating separation and partial analyte characterization. In order to harness the intrinsic analytical potential of the capillary electrophoretic process, its theoretical underpinnings have been investigated widely, \cite{Reijenga, Dagan} and references therein. \\

The simulation and modeling of many electrophoretic processes is complex, and often only  a numerical solution to the underlying partial differential equation is accessible. Despite being extremely useful analytical tools in many cases, they however also generally inhibit the possibility for parameter estimation due to the absence of a physicochemical model. In this letter, we introduce a fast, versatile and reliable C\verb!++! tool for parameter estimation using nonlinear least-squares optimization \cite{Nocedal}. We model the electrophoretic process using a specific closed-form solution to the Nernst-Planck equation. This approach facilitates physicochemical parameter estimation of all, potentially many dozens, detected molecular species in a complex mixture. \\

The electrophoretic process of a molecular species $i$ in a capillary filled with an electrolyte is modeled by the solution $c_{i}(x,t)$ of the one-dimensional convection-diffusion equation
\begin{equation}
	\frac{\partial c_{i}}{\partial t} = D_{i}\frac{\partial^{2} c_{i}}{\partial x^{2} } - v_{i}\frac{\partial c_{i}}{\partial x }
\end{equation}
Here, $c_{i}(x,t)$ denotes the one-dimensional relative concentration of species $i$ with respect to space $x$ and time $t$. $D_{i}$ represents the diffusion coefficient and $v_{i}$ the electrophoretic velocity.

For a complex mixture of $n$ detected molecular species subject to separation by capillary zone electrophoresis, the entire electropherogram is modeled as  
\begin{equation}
	M(x,t) = \sum_{i=1}^{n} c_{i}(x,t)
\end{equation}
The program finds the model $M^{*}(x,t)$ that best approximates the data. In particular, 
it finds the best estimates over finite intervals for $v$, $D$ and $c_{0}$ for all $n$ detected species.

\section{Operation and Features}
In order to apply \textit{awkt} successfully, the CZE experiment must be conducted adhering to some well-defined experimental requirements. In particular, the initial width of the injected zone $w_{0}$, the distance from the capillary entrance to the detector location $L_{d}$, the electric field strength $E$, and the system temperature $T$ must be known to some acceptable level of accuracy. In fact, it is critical for the analysis, that the experiment is conducted in strictly isothermal conditions. Otherwise, the peak broadening characteristics cannot conceivably be anticipated with the currently employed model. However, the current model is amenable to modification. Any corrections to the current model may be added in the source code without affecting the overall structure of the program. Furthermore, the current model can be completely replaced by another, potentially more complex model. It is also imperative to ensure that the background electrolyte concentration remains constant throughout the experiment. \\

The program has a number of user-adjustable data processing and analysis options. A list of the currently available options is shown below. For details please see the program documentation \cite{Loeffeld}.\\

\begin{itemize}[rightmargin=-0.3mm]
	\item peak detection
	\item hypothesis testing 
	\item model selection
	\item single file and batch processing 
	\item concurrent batch processing
	\item time-range selection
	\item noise filtering (gaussian kernel smoothing)
	\item data  and model visualization 
	\item global baseline rectification 
	\item local baseline rectification 
	\item local baseline estimation\\ 
\end{itemize}

We give a brief high-level overview of the program operations in the default setting. The program loads the data and its associated experimental conditions from two separate files. A procedure attempts to detect as many peaks as possible given a preset detection sensitivity. Another procedure then, splits the data into frames that each contain at least one but potentially multiple peaks, again depending on a preset upper bound and peak separation tolerance. The objective is to have the fewest number of peaks possible in a frame because it vastly simplifies and speeds up the optimization process. Subsequently, given the supplied model derived from the Nernst-Planck equation, or any other supplied model for that matter, the data in the identified data frames are subject to nonlinear least-squares optimization. Since the peak detection procedure may have missed real peaks, we enable another algorithm to vary the number of detected peaks in order to potentially find a better model to the data, however only accept the new model if the value of the associated objective function has decreased sufficiently. Finally, for each identified data frame we use its best model, i.e. the best estimates for the diffusion coefficients and electrophoretic velocities, together with the experimental conditions and the Stokes-Einstein relations, to compute the hydrodynamic radii and the net electric charges associated to the frames.
\section{Software Requirements}
\textit{awkt} is built on top of the \textit{Ceres Solver} \cite{Agarwal}, an open source C\verb!++! library for modeling and solving large, complicated optimization problems. Another crucial component is the C\verb!++! peak detection library, \textit{Persistence1d} \cite{Kozlov}. For visualization of the data and the computed models, the \textit{Gnuplot-i} C\verb!++! interface \cite{Devillard} is included. The \textit{Boost} C\verb!++! library \cite{boost} is used for file and directory management. All other features are due to the C\verb!++! Standard Template Library. For installation details, we refer to the documentation of \textit{awkt} \cite{Loeffeld}.

\section{Conclusion}
We introduce \textit{awkt}, a fast, versatile and robust  C\verb!++! parameter estimation tool aimed to assist molecule characterization efforts using capillary zone electrophoresis. The program utilizes the electropherogram obtained from a CZE experiment to estimate numerical values for the diffusion coefficient, hydrodynamic radius and the net electric charge of all molecular entities that are detected during the EPG analysis process. As a result, numerical estimates for any associated physical quantity, such as the molecular weight for instance, may also be obtained given a mathematical relation between the quantities is available or can be established \cite{Erickson}.

\end{document}